\documentclass[a4paper, twocolumn]{revtex4}

\usepackage{rotating}
\usepackage{epstopdf}
\usepackage{color}
\usepackage{amsmath}
\usepackage{graphicx}
\usepackage{hyperref}
\usepackage{color}
\usepackage{amssymb}
\usepackage{dcolumn}
\usepackage{bm}
\usepackage{epsfig}
\usepackage{array}
\usepackage{subfigure}
\usepackage{upgreek}
\usepackage{wasysym}
\usepackage{slashed}
\usepackage[section]{placeins}
\usepackage{appendix}
\usepackage{subfigure}
\usepackage{adjustbox}
\newcommand{\be}{\begin{equation}} 
\newcommand{\ee}{\end{equation}} 
\newcommand{\bea}{\begin{eqnarray}} 
\newcommand{\eea}{\end{eqnarray}} 
\newcommand{\bqa}{\begin{eqnarray}}
\newcommand{\eqa}{\end{eqnarray}}

\newcommand{\figpot}{
\begin{figure}[h]
\centering
\hspace{0cm}
\includegraphics[height=35mm,width=70mm]{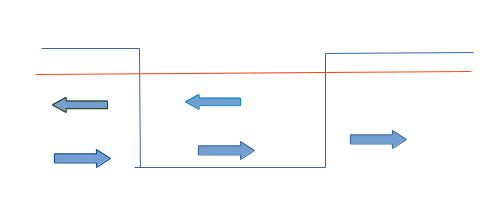}
\caption{Schematic diagram of a potential well chosen in our case. Here the solid blue arrows indicates the incident and reflected electrons in the left and center region, while there is no reflection in the right region. Red line corresponds to the Fermi energy level to confined the electrons in the well. Here we are mainly concerned about the center region.}
\label{fig:pot}
\end{figure}
}

\newcommand{\figTCo}{
\begin{figure}[h]
\centering
\hspace{0cm}
\subfigure[noonleline][]
{\label{fig:TCo4}\includegraphics[height=35mm,width=40mm]{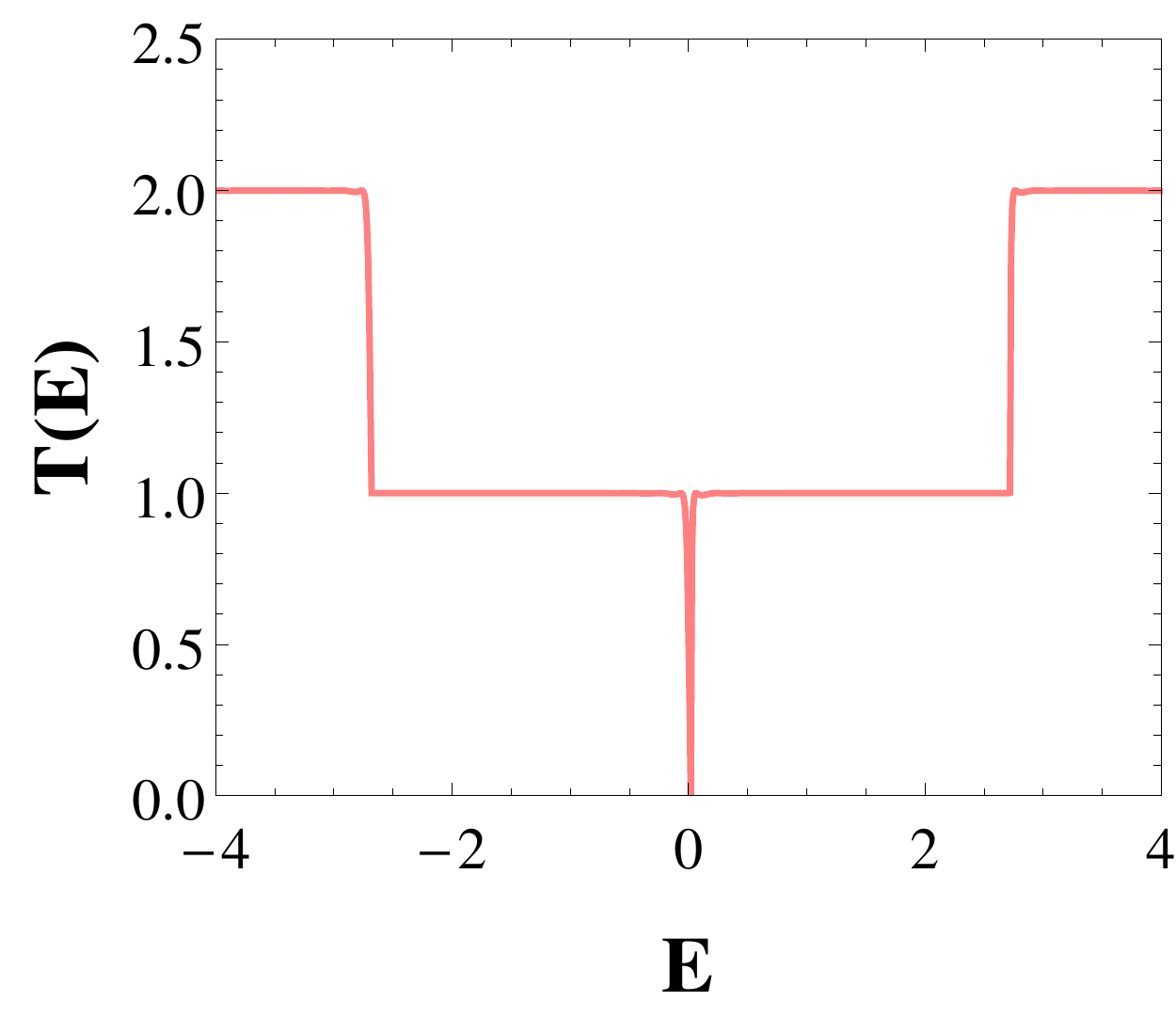}}
\hspace{0cm}
\subfigure[noonleline][]
{\label{fig:TCo16}\includegraphics[height=35mm,width=40mm]{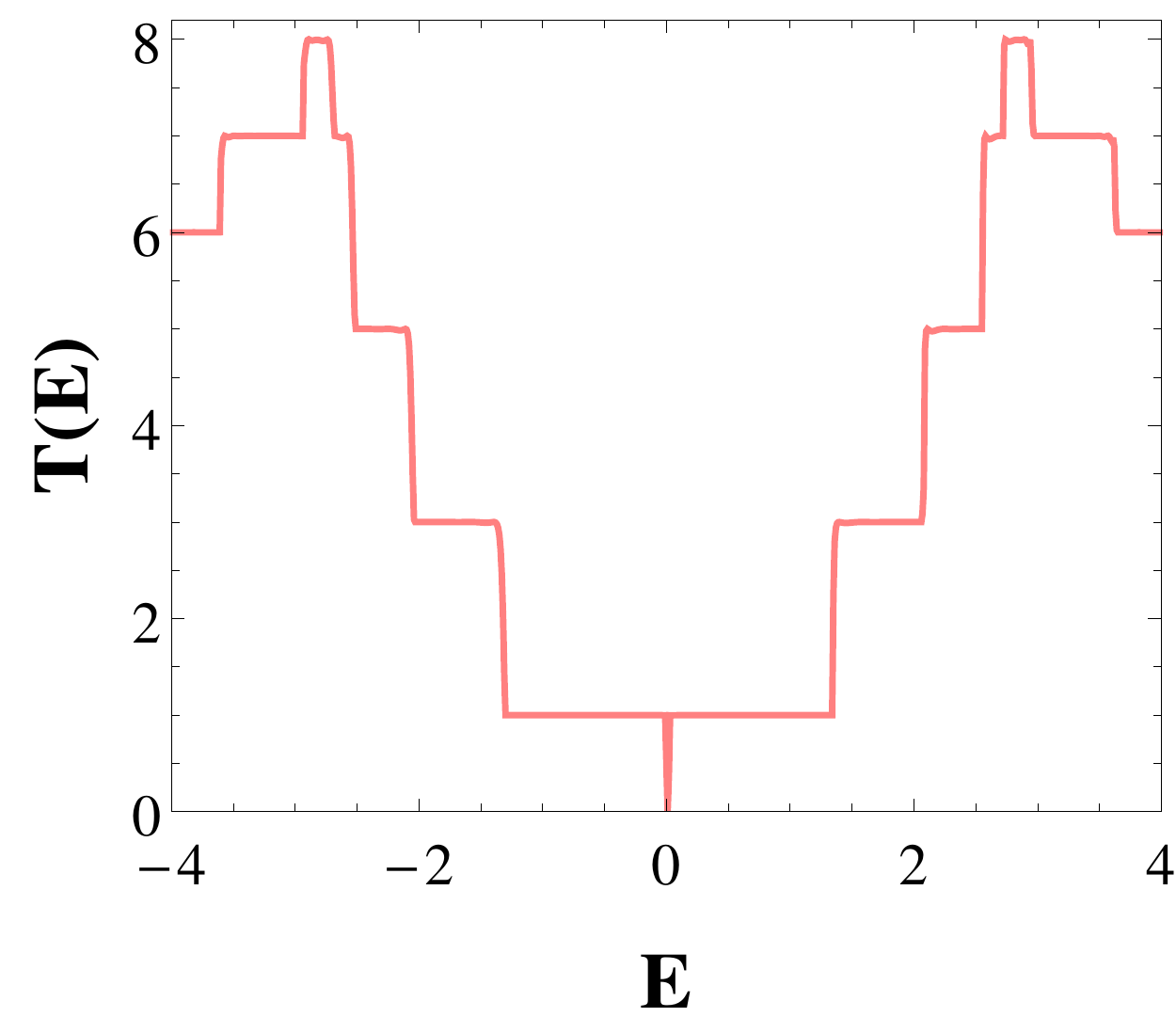}}
\subfigure[noonleline][]
{\label{fig:TCo64}\includegraphics[height=35mm,width=40mm]{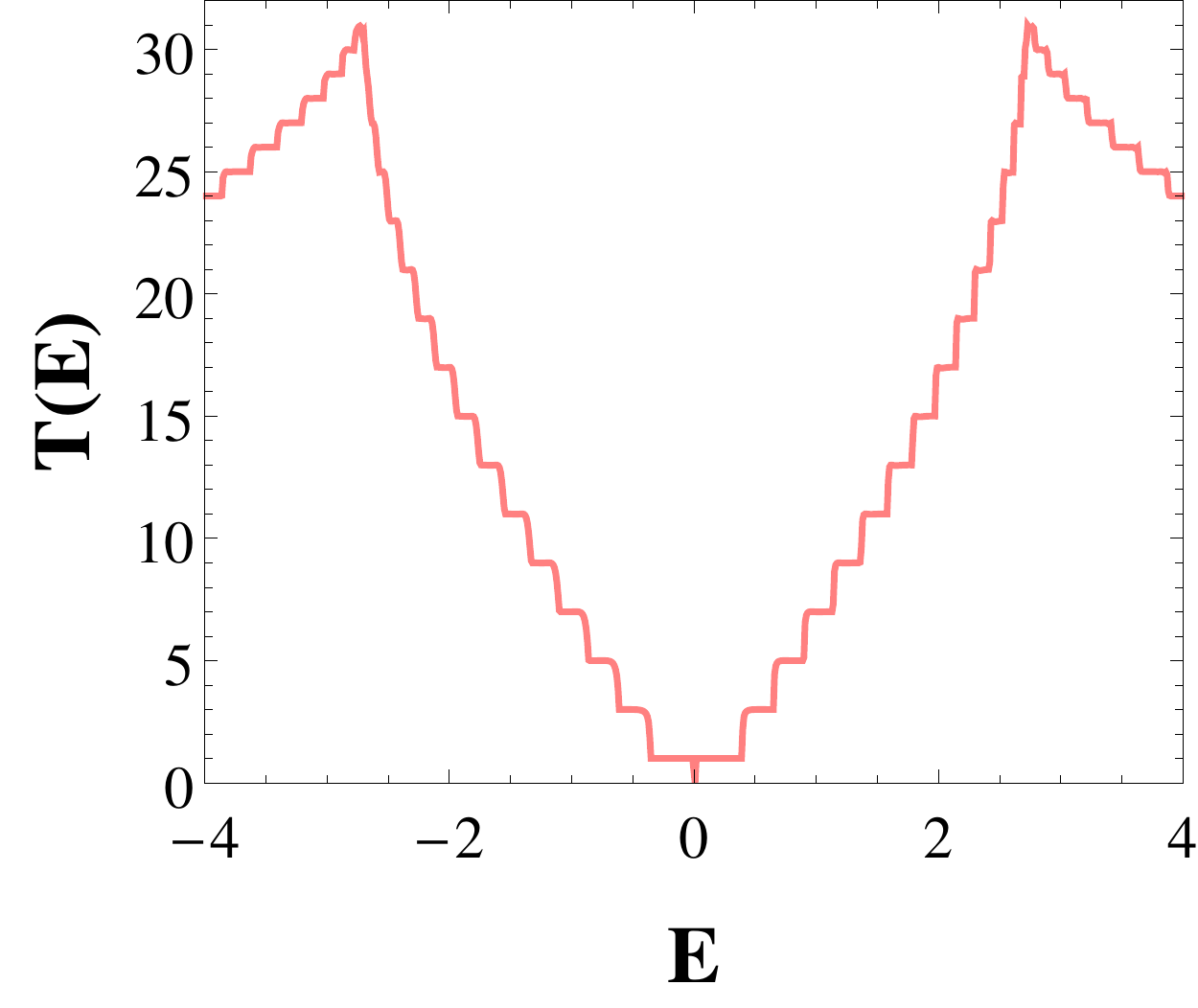}}
\caption{Energy dependent transmission coefficient is shown for different number of atoms per unit cell (a): N=4; (b): N=16; and (c): N=64.}
\label{fig:TCo}
\end{figure}
}
\newcommand{\figBS}{
\begin{figure}[h]
\centering
\hspace{0cm}
\subfigure[noonleline][]
{\label{fig:BS4}\includegraphics[height=35mm,width=40mm]{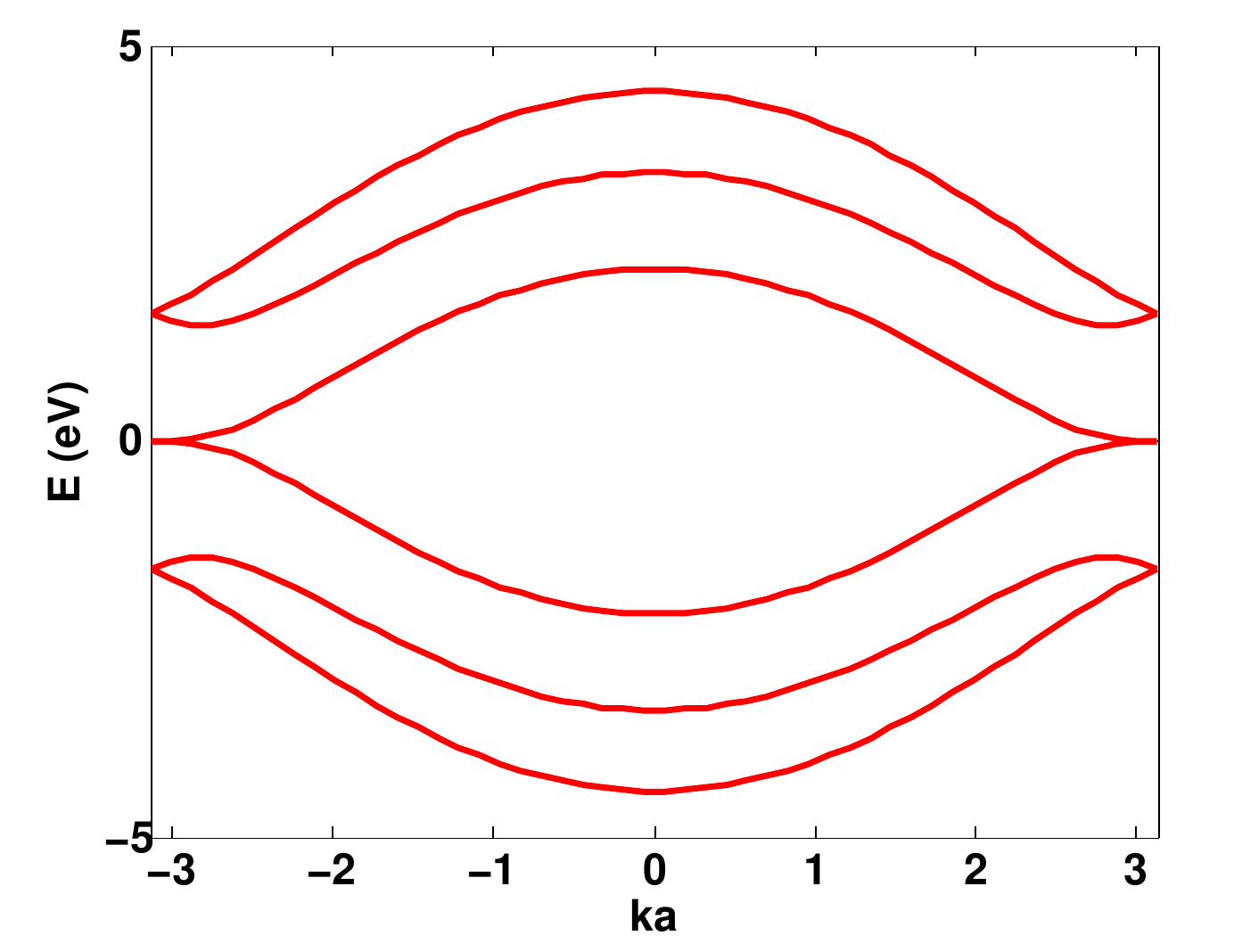}}
\hspace{0cm}
\subfigure[noonleline][]
{\label{fig:BS6}\includegraphics[height=35mm,width=40mm]{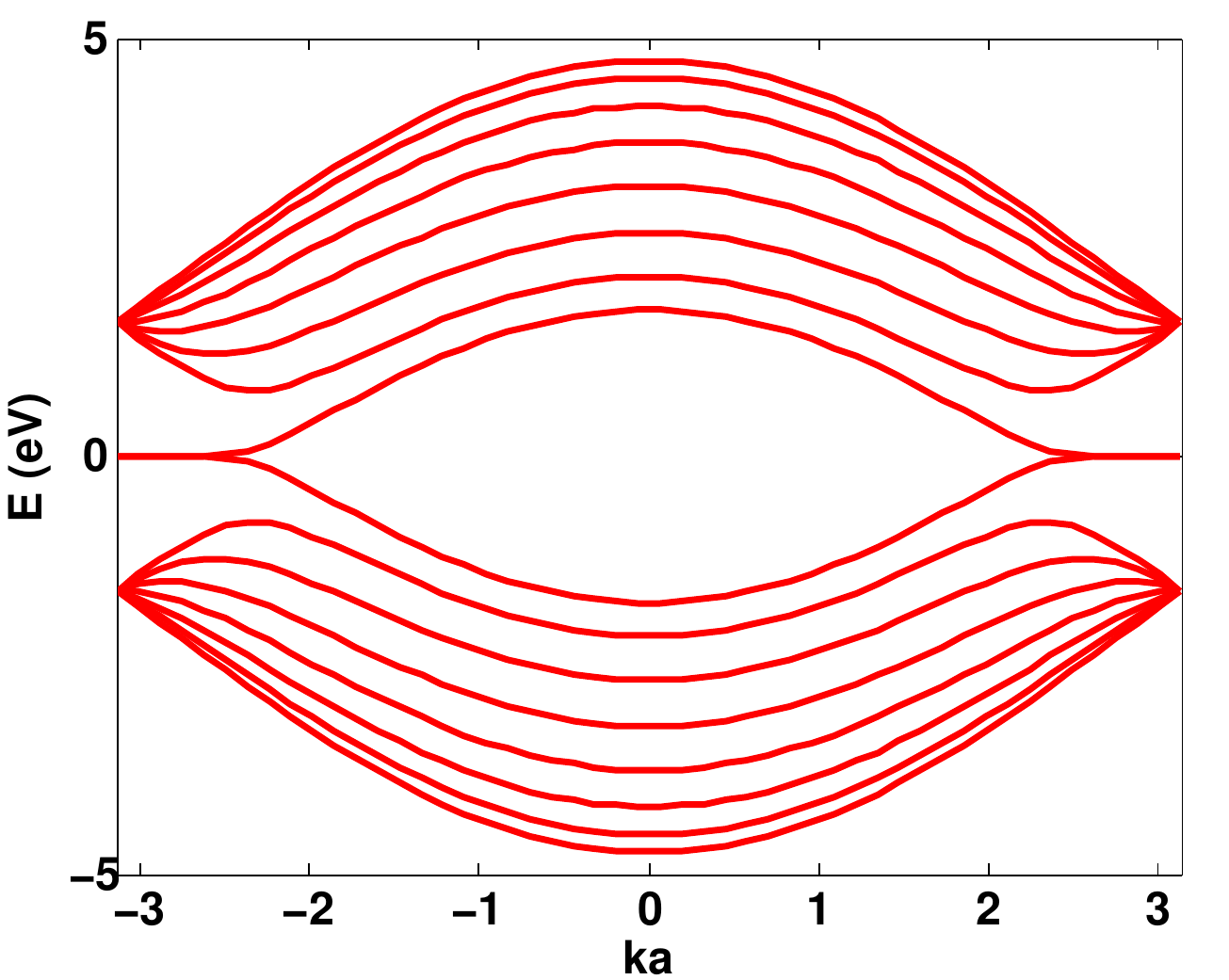}}
\subfigure[noonleline][]
{\label{fig:BS4}\includegraphics[height=35mm,width=40mm]{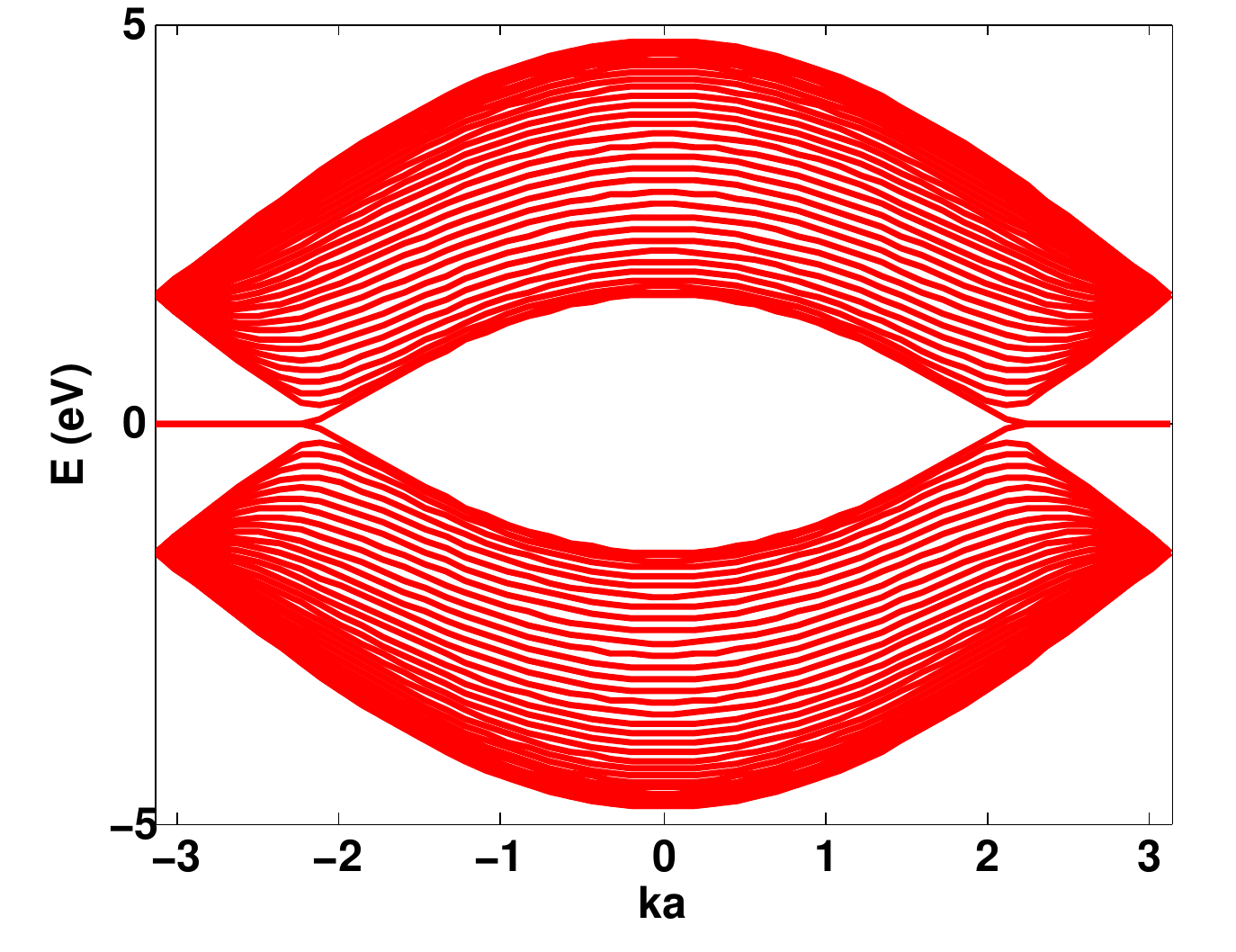}}
\caption{Plot of band structure corresponds to (a) N=6; (b): N=16; and (c): N=64.}
\label{fig:BS}
\end{figure}
}

\newcommand{\figLDOS}{
\begin{figure}[h]
\centering
\hspace{0cm}
\subfigure[noonleline][]
{\label{fig:LDO4}\includegraphics[height=35mm,width=40mm]{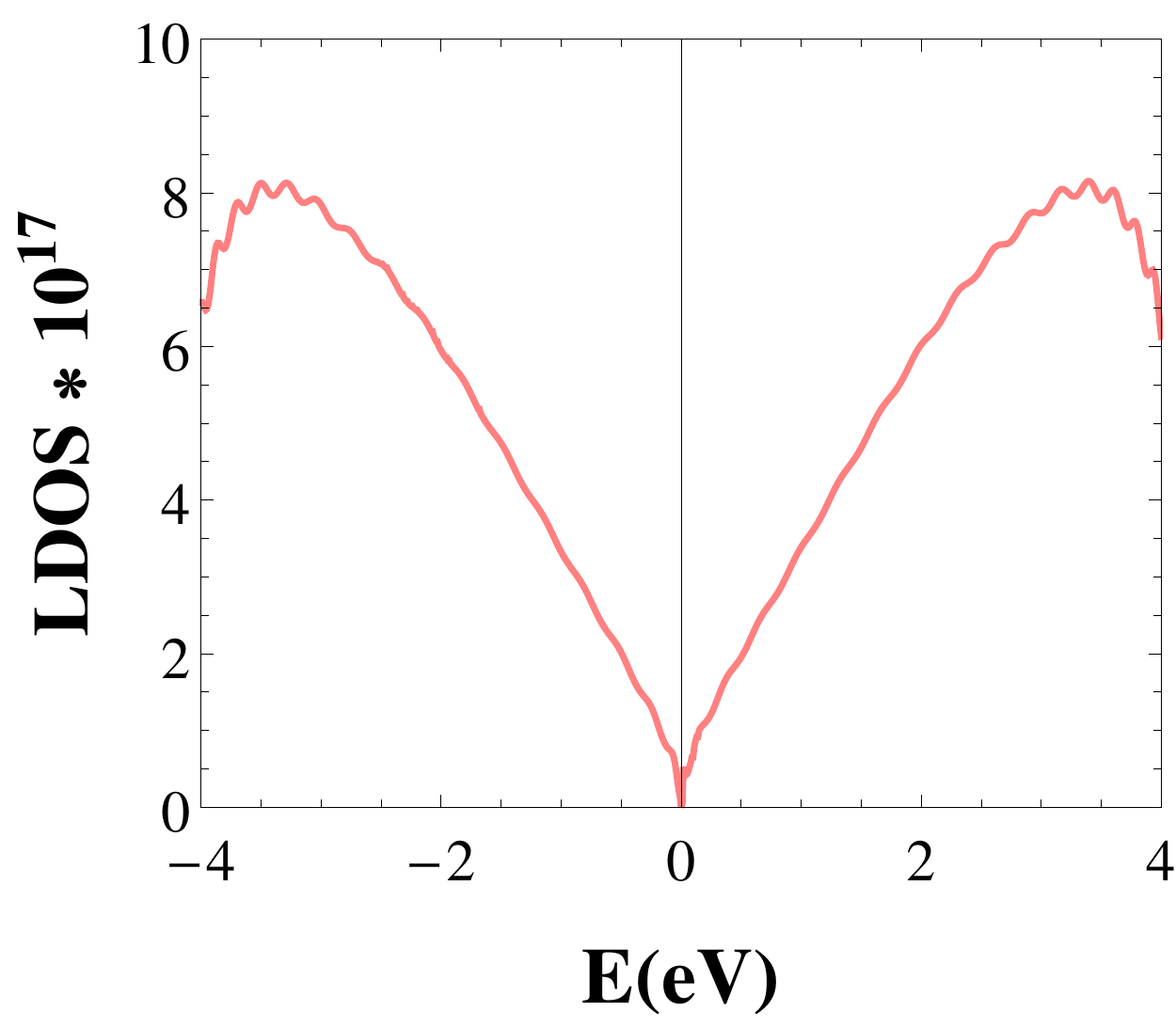}}
\hspace{0cm}
\subfigure[noonleline][]
{\label{fig:LDO16}\includegraphics[height=35mm,width=40mm]{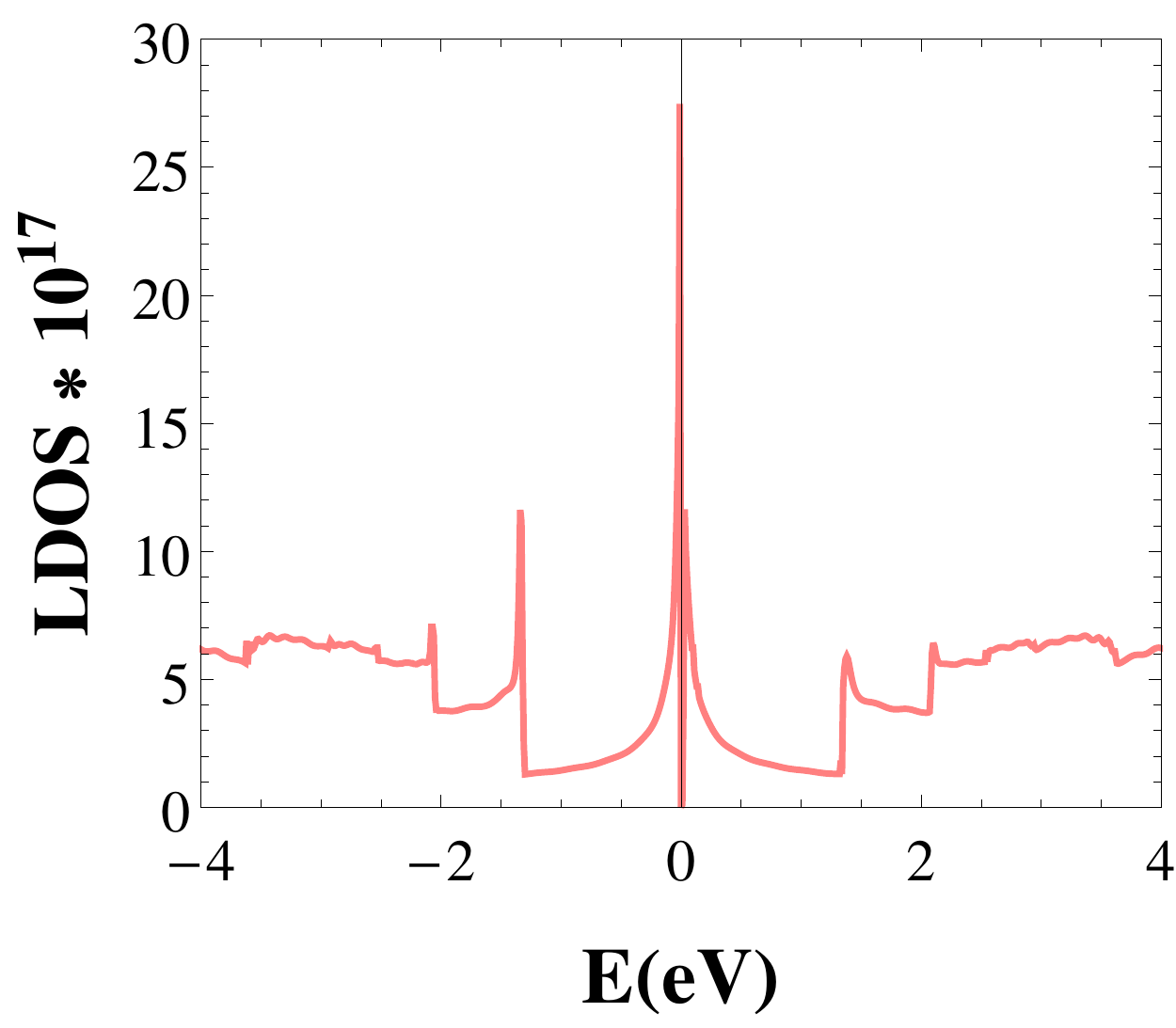}}
\subfigure[noonleline][]
{\label{fig:LDO64}\includegraphics[height=35mm,width=40mm]{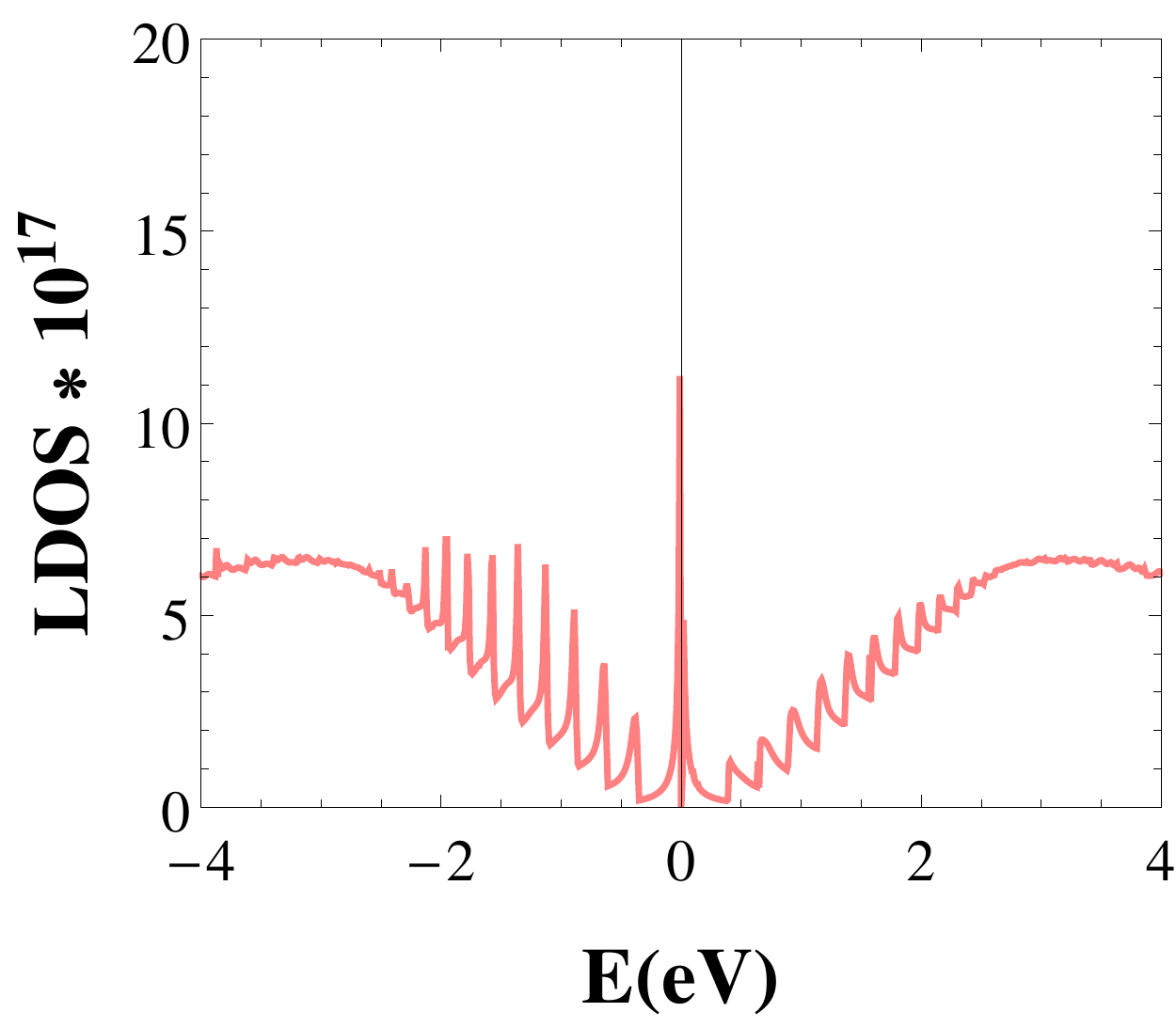}}
\caption{The local density of states (LDOS) is shown for different number of atoms per unit cell (a): N=4; (b): N=16; and (c): N=64.}
\label{fig:LDOS}
\end{figure}
}

\newcommand{\figEC}{
\begin{figure}[h]
\centering
\hspace{0cm}
\subfigure[noonleline][]
{\label{fig:EC4}\includegraphics[height=35mm,width=40mm]{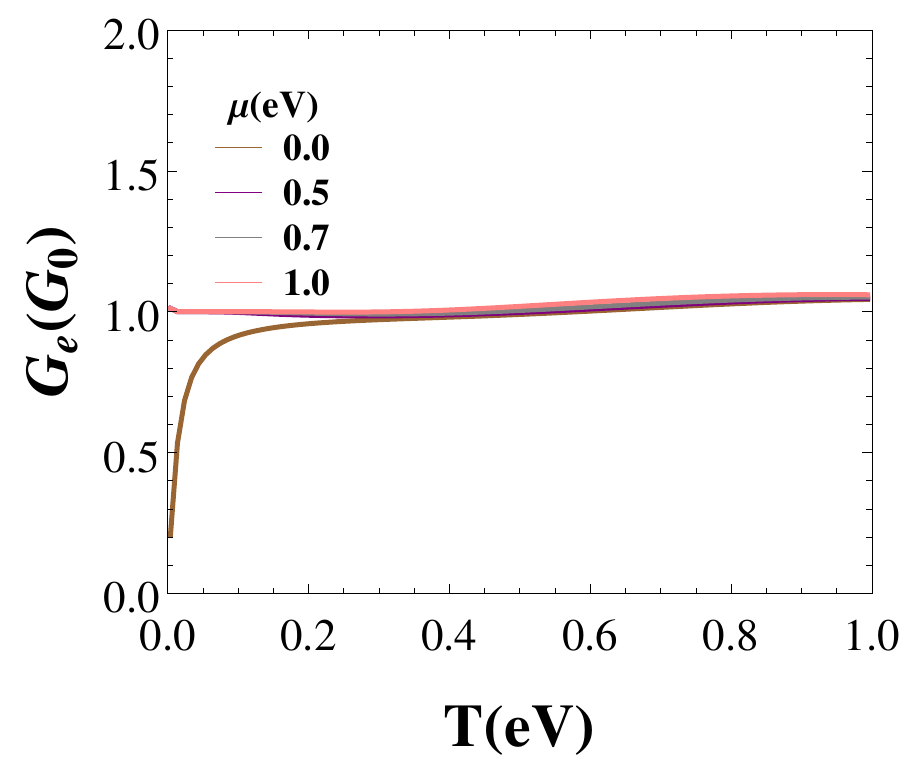}}
\hspace{0cm}
\subfigure[noonleline][]
{\label{fig:EC16}\includegraphics[height=35mm,width=40mm]{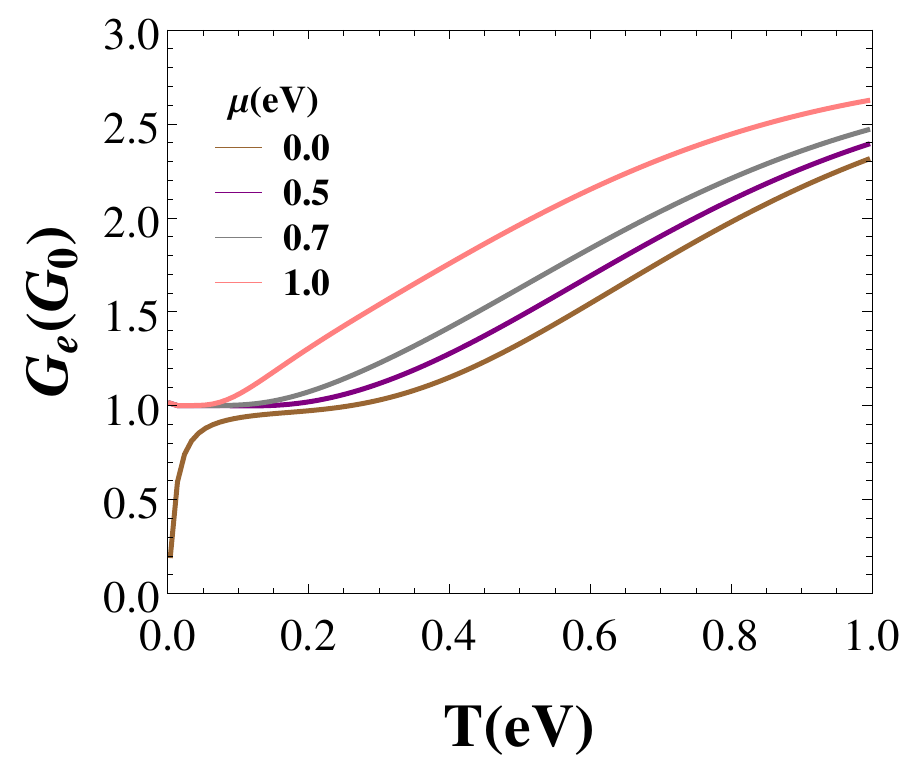}}
\subfigure[noonleline][]
{\label{fig:EC64}\includegraphics[height=35mm,width=40mm]{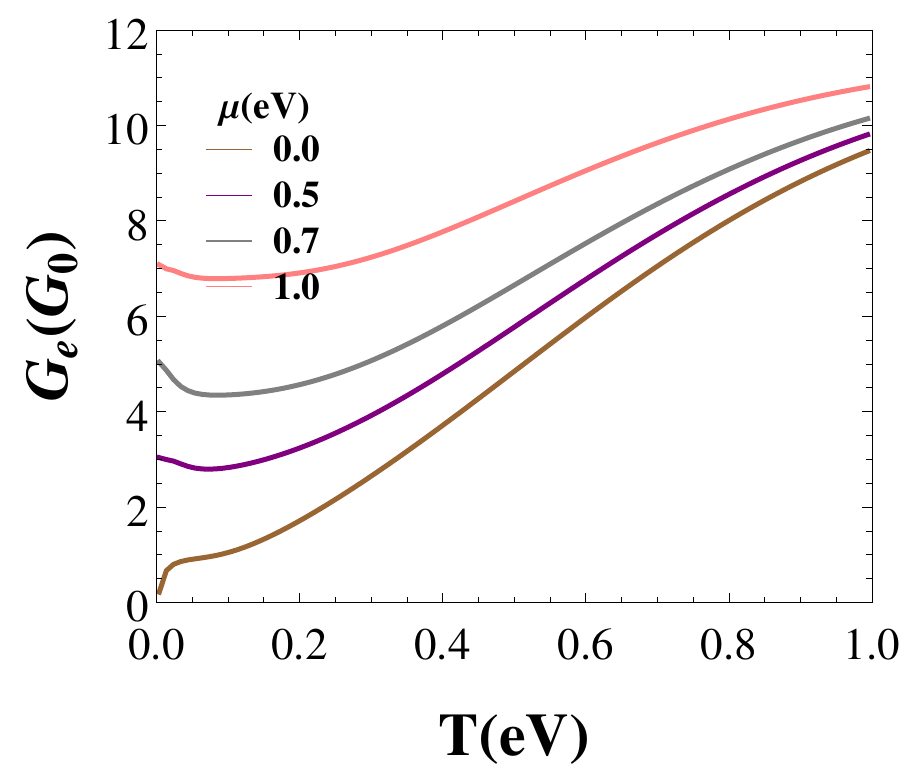}}
\caption{Variation of the electron conductance as a function of temperature $T$(eV) for different number of atoms per unit cell (a): N=4; (b): N=16; and (c): N=64. Here $G$ is measured in terms of $G_{0}= 2e^{2}/h$.}
\label{fig:EC}
\end{figure}
}

\newcommand{\figTC}{
\begin{figure}[h]
\centering
\hspace{0cm}
\subfigure[noonleline][]
{\label{fig:TC4}\includegraphics[height=35mm,width=40mm]{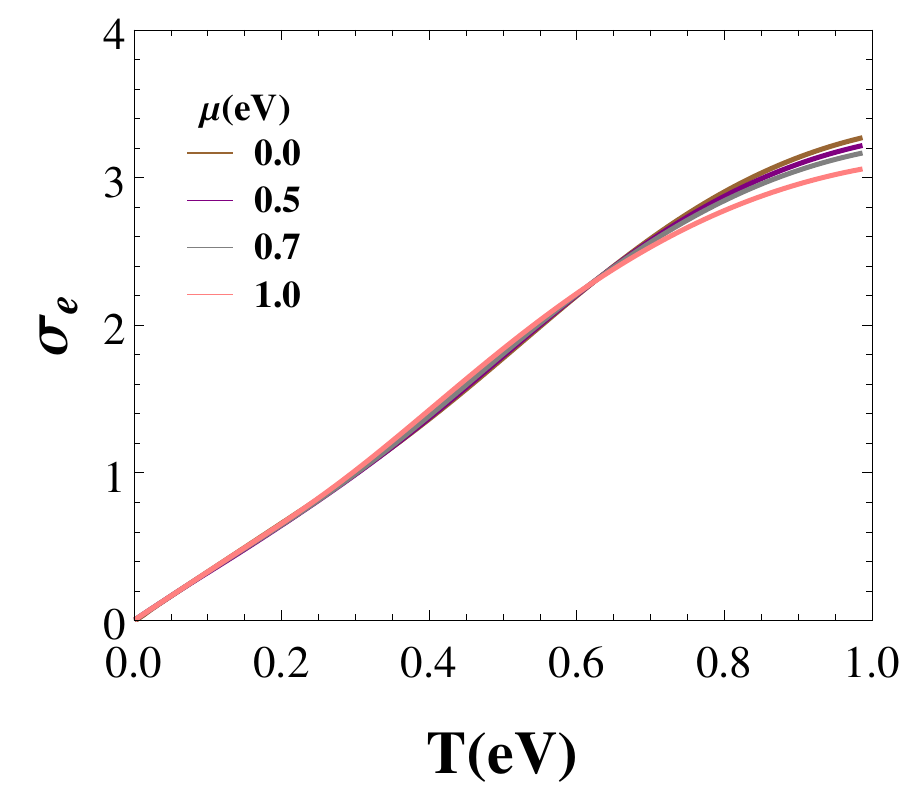}}
\hspace{0cm}
\subfigure[noonleline][]
{\label{fig:TC16}\includegraphics[height=35mm,width=40mm]{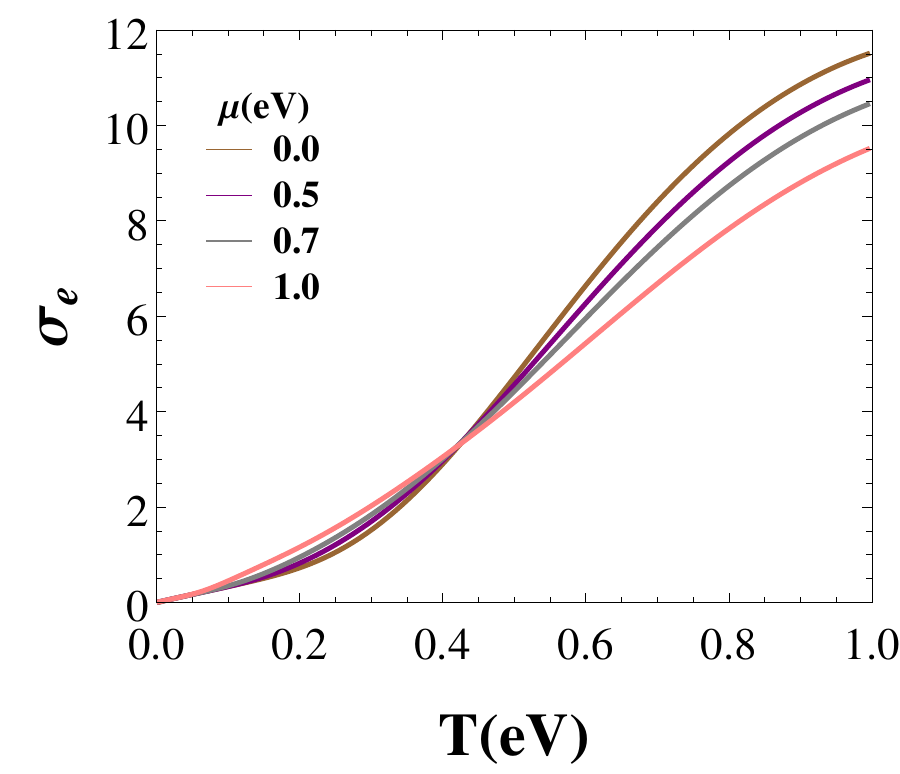}}
\subfigure[noonleline][]
{\label{fig:TC64}\includegraphics[height=35mm,width=40mm]{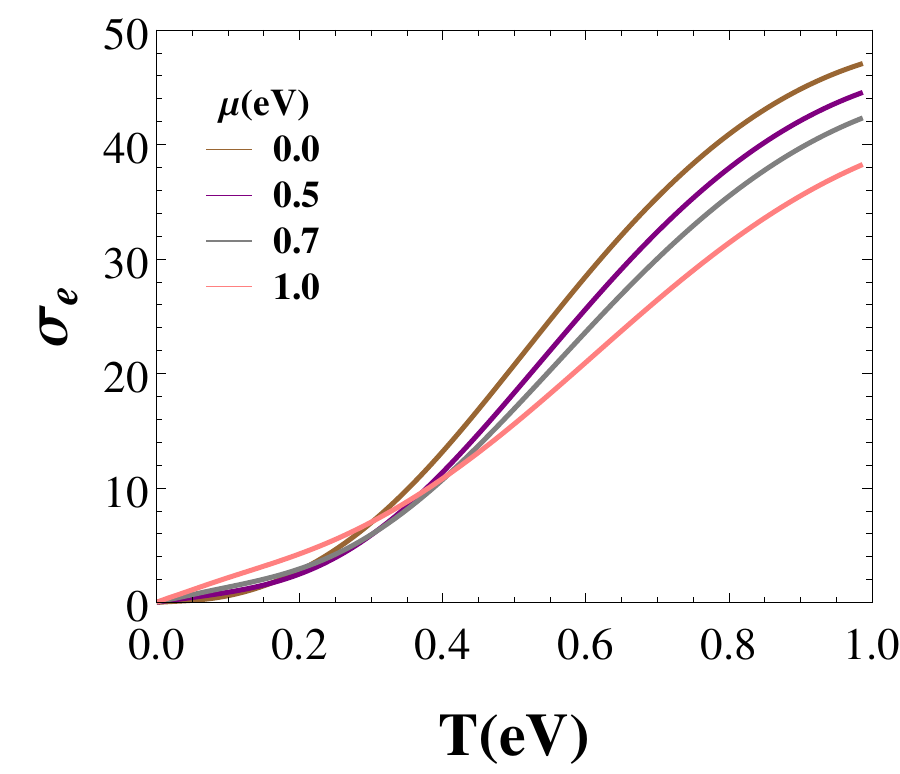}}
\caption{Variation of the thermal conductance as a function of temperature $T$(eV) for different number of atoms per unit cell (a): N=4; (b): N=16; and (c): N=64.}
\label{fig:TC}
\end{figure}
}

\newcommand{\figSC}{
\begin{figure}[h]
\centering
\hspace{0cm}
\subfigure[noonleline][]
{\label{fig:SC4}\includegraphics[height=35mm,width=40mm]{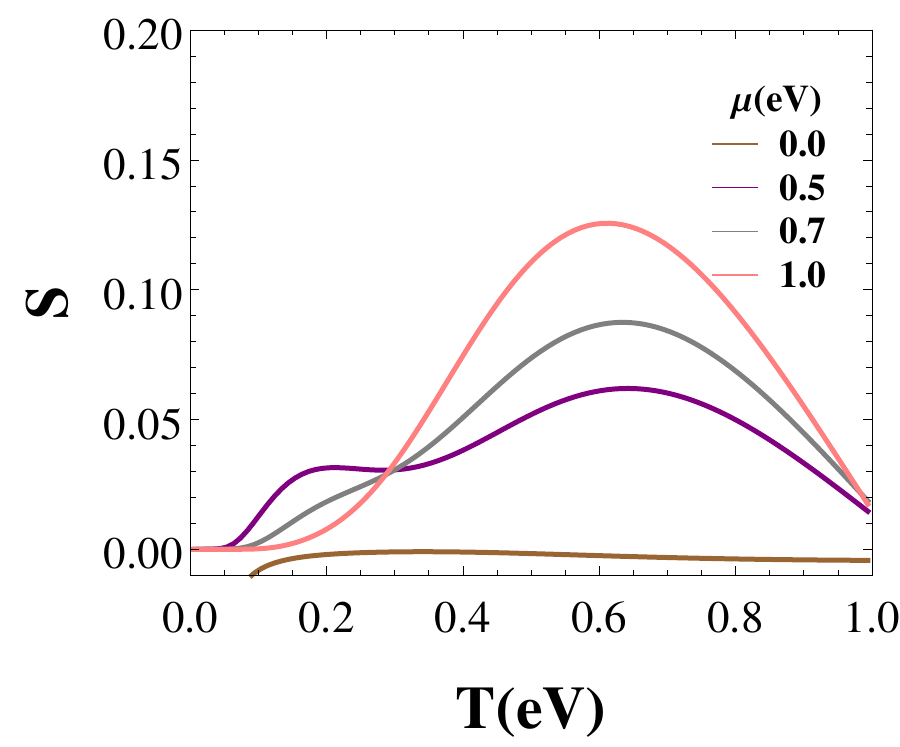}}
\hspace{0cm}
\subfigure[noonleline][]
{\label{fig:SC16}\includegraphics[height=35mm,width=40mm]{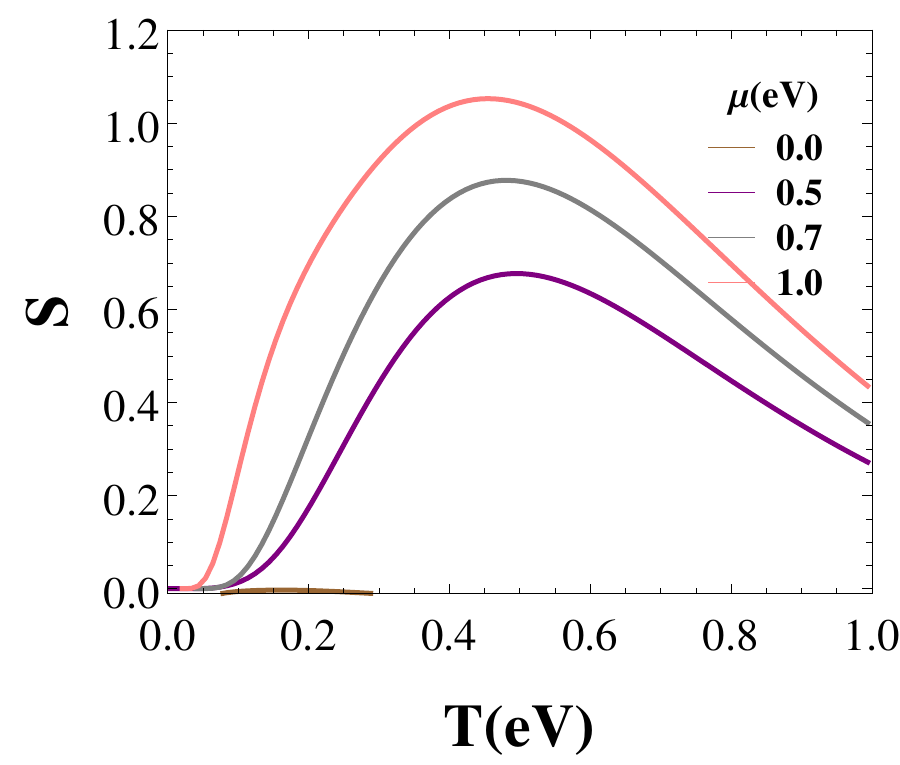}}
\subfigure[noonleline][]
{\label{fig:SC64}\includegraphics[height=35mm,width=40mm]{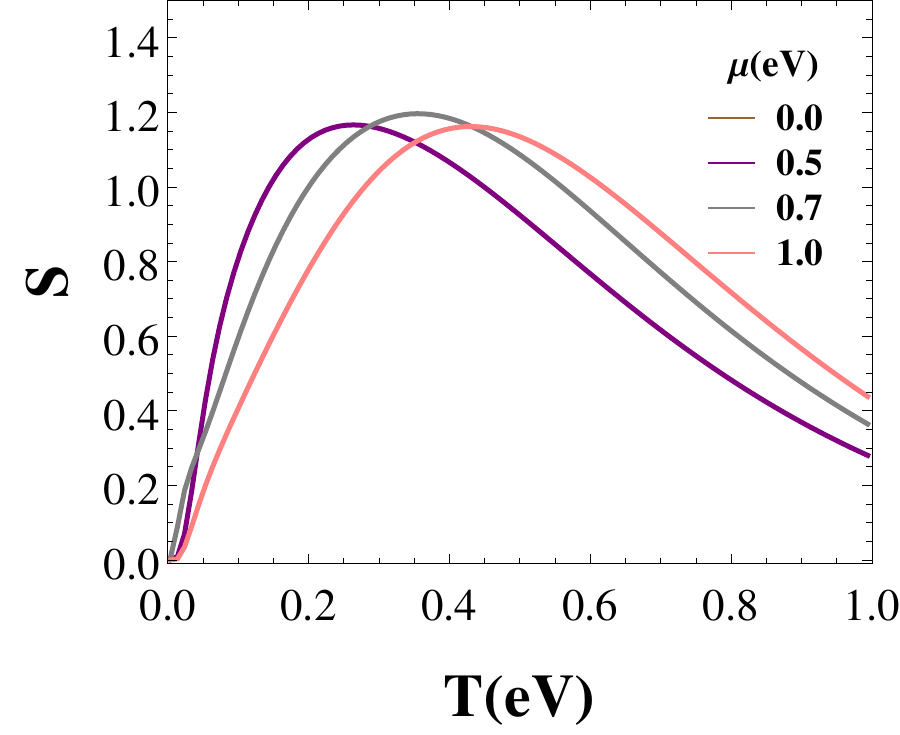}}
\caption{The Seebeck coefficient as a function of temperature $T$(eV) with different number of atoms per unit cell for ZGNR (a): N=4; (b): N=16; and (c): N=64.}
\label{fig:SC}
\end{figure}
}

\newcommand{\figFM}{
\begin{figure}[h]
\centering
\hspace{0cm}
\subfigure[noonleline][]
{\label{fig:FM4}\includegraphics[height=35mm,width=40mm]{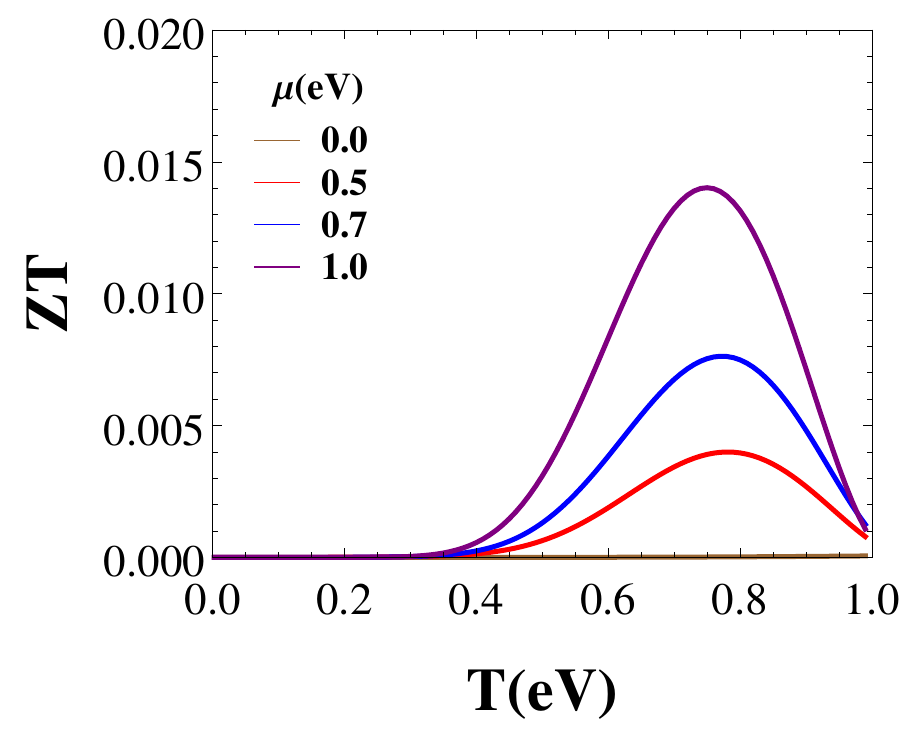}}
\hspace{0cm}
\subfigure[noonleline][]
{\label{fig:FM16}\includegraphics[height=35mm,width=40mm]{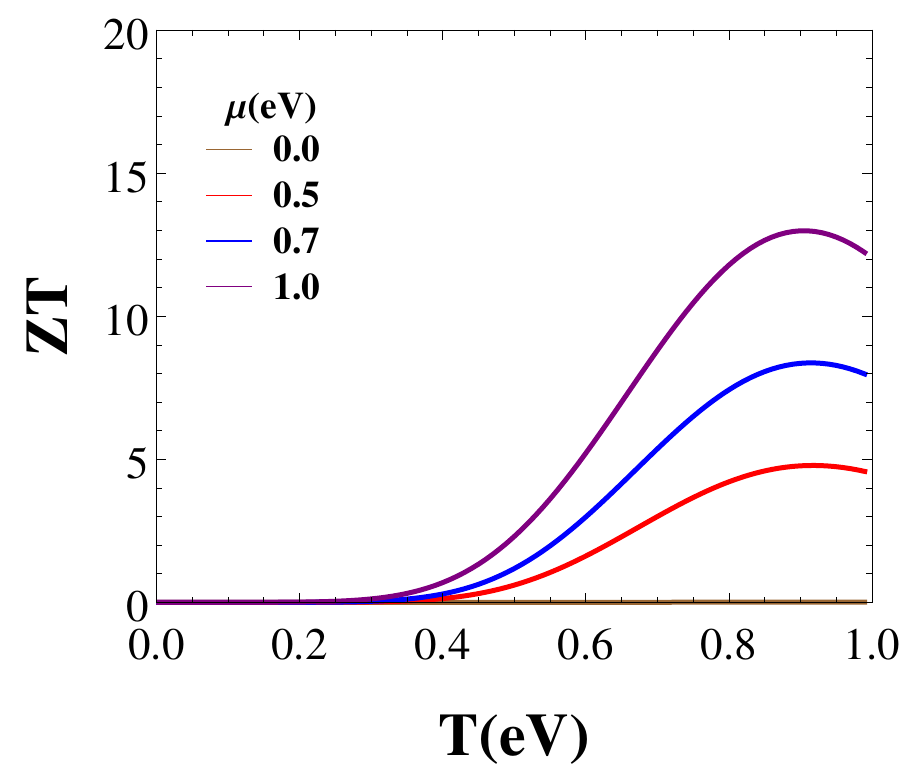}}
\subfigure[noonleline][]
{\label{fig:FM64}\includegraphics[height=35mm,width=40mm]{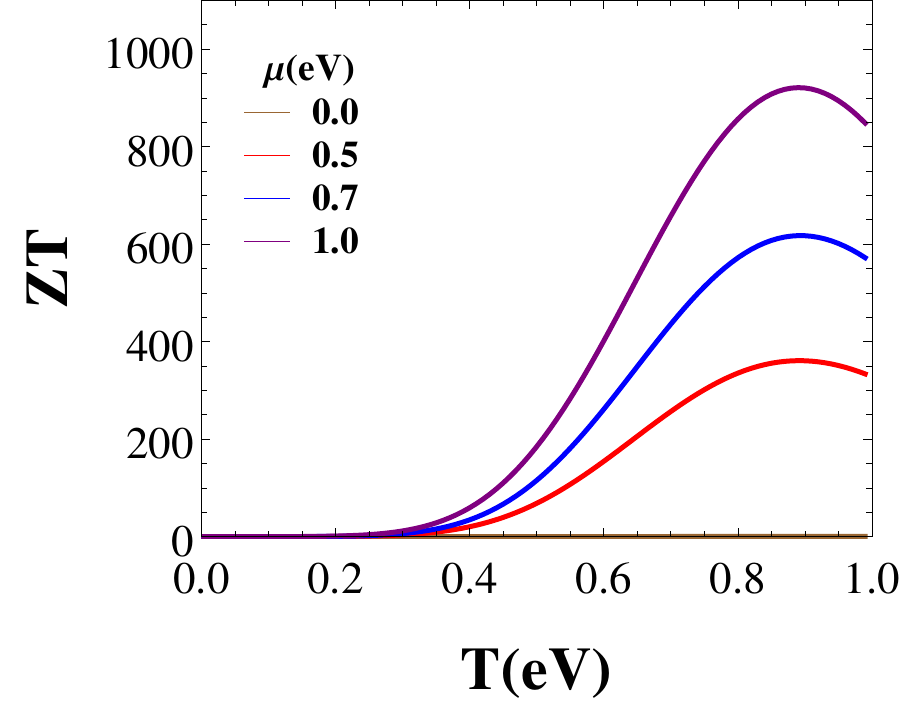}}
\caption{Plot for the Figure of merit as a function of temperature $T$(eV) with different number of atoms per unit cell for ZGNR (a): N=4; (b): N=16; and (c): N=64 and at different chemical potential values.}
\label{fig:FM}
\end{figure}
}
\begin{document}
\title{Aspects of electron transport in zigzag graphene nanoribbons}

\author{Pankaj Bhalla}
\email{pankajbhalla66@gmail.com}
\affiliation{Physical Research Laboratory, Navrangpura, Ahmedabad-380009, India.}

\author{Surender Pratap}
\email{suren1986dhalaria@gmail.com}
\affiliation{Department of Physics, BITS Pilani, Pilani campus-333031(Rajasthan) India.}



\begin{abstract}
We investigate the aspects of the electron transport in the zigzag graphene nanoribbons (ZGNRs) using the non-equilibrium Green's function (NEGF) formalism. The latter is an esoteric tool in mesoscopic physics and using this tool the analysis is performed by considering the potential well within the system. Within this potential well, the dependence of the transmission coefficient, local density of states (LDOS) and the electron transport properties on the number of atoms per unit cell is discussed. We observe that the electron and thermal conductance increases with the increase of the number of atoms. In addition to these, the figure of merit which depends on these transport properties has also been studied and we find that the electrons' contribution to enhance the figure of merit is important above the crossover temperature. 
\end{abstract}
\maketitle
Keywords: Graphene, Zigzag graphene nanoribbons, Transport
\section{Introduction}
\label{sec:intro}
The discovery of graphene, a two dimensional material has created an immense interest in the fundamental physics and device research communities\cite{novoselov_04}, \cite{geim_07}, \cite{sarma_07}, \cite{geim_09}. This interest is continued growing due to its unique properties such as linear dispersion relation, unusual quantum Hall effect, thermal and electrical transport\cite{kane_05}, \cite{son_06a}, \cite{rycerz_07}, \cite{bhalla_17}. Graphene can also be cast into quasi one dimensional nanoribbons which makes the study of this material more fascinating. The latter are thin strips of graphene having width smaller than $50$nm\cite{acik_11}, \cite{dutta_10} and obtained by cutting the infinite two dimensional sheets in different ways which leads to various edge structures. Depending on the edge structures, the graphene nanoribbons (GNRs) can be characterized as zigzag and armchair GNRs\cite{wakabayashi_99}. The armchair GNRs are those which have no edge 
reconstruction and the zigzag GNRs have edge states\cite{nakada_96}, \cite{fujita_96}, \cite{barone_06}, \cite{ezawa_06}, \cite{son_06}, \cite{orlof_13}. The latter feature creates more attention to researchers to study ZGNRs. Also, the localized edge states which occurs only in case of ZGNRs lead to the magnetic ordering\cite{yazyev_08}, \cite{yazyev_10},\cite{magda_14} and renders this material as a suitable candidate for the technological advancements such as spintronics\cite{zhang_14}. 

The electronic transport properties of ZGNRs show number of interesting phenomenon due to their edge states. Although these transport properties have been examined theoretically and experimentally\cite{wakabayashi_12}, \cite{tong_book},\cite{li_12}. However, the detailed aspects of the electron transport is lacking. To perform this, the detailed analysis of the transport properties such as electron conductance, thermal conductance and thermoelectric coefficient of ZGNRs from microscopic point of view is very important. Furthermore, this analysis is also significant as it deals with the figure of merit which is an extremely important for the application purposes\cite{nolas_book}, \cite{kamran_book}. However, the figure of merit depends upon the electrical conductivity, thermal conductivity and Seebeck coefficient\cite{kamran_book}. The latter properties depend on the transmission of the charge carriers through the potential barrier or the excitation of electrons from valence to the conduction bands. In this context, several works have been done to enhance the figure of merit of the material by studying most of the thermoelectric properties\cite{mahan_97}, \cite{chen_03},\cite{nolas_06}, \cite{dresselhaus_07}. The major challenge of this work depends on the relationship between the transport properties as discussed earlier. It is difficult to study figure of merit based on the modified behavior of one of the property and keeping other remain unaltered. It has been studied in the bismuth and tellurium compounds, carbon single walled nanotubes\cite{chung_97}, \cite{chung_00}, \cite{jiang_11}, \cite{balandin_10}. Here it is argued that the performance of the thermoelectric materials can be improved with the application of quantum well structures for the confinement of electrons\cite{hicks_93}, \cite{hyldgaard_97}, \cite{chen_98}, \cite{venkat_00}, \cite{lin_03}, \cite{yang_08}. In addition to these, it is also studied with the geometry and size effects, surface roughness methods, etc\cite{shi_09}, \cite{liang_10},\cite{ni_08},\cite{mohiuddin_09}, \cite{huang_09}, \cite{xu_09}.

In his pathway, we have done calculation here by using the non-equilibrium Green's function (NEGF) approach\cite{datta_book}. These properties have been investigated by solving the transport equations for the electrons and phonons and found that the quasi one dimensional structure is an important to understand the thermoelectric transport properties of graphene nanoribbons\cite{datta_book}, \cite{hwang_09}, \cite{liang_09}, \cite{ouyang_09}, \cite{haug_book}. The more emphasize is given to study the effects of edge roughness, vacancy defects, strain effects\cite{martin_09}, \cite{donadio_09}, \cite{ni_08}, \cite{mohiuddin_09}, \cite{huang_09}, \cite{xu_09}. However, the investigation of these transport properties based on the variation of the number of atoms per unit cell in a confined well is not addressed in detail. This is an important aspect as it effects directly the transmission coefficient of the system.

In this scenario, efforts have been given to study the phonon transport\cite{wang_08}, \cite{datta_book}, \cite{haug_book} more and less emphasize is given to the electron transport. However to compare the results with experimental facts, it is also equally important to study the electron transport which is the main aim of this paper. In this view, we investigate the transmission coefficient and the electron transport properties such as electron conductance, thermal conductance and Seebeck coefficient with the variation of the number of atoms per unit cell and with the change in the chemical potential. We find due to the peak occurs in the local density of states near the Fermi energy, the electron and the thermal conductance increases with the increase in the number of atoms\cite{fujita_96}, \cite{nakada_96}, \cite{wakabayashi_99}. 

This paper is organized as follows: Firstly, the theoretical background to study the transport is discussed in Sec.~\ref{sec:results}. Here we discussed the model Hamiltonian Sec.~\ref{sec:model} and basics of the non-equilibrium Green's function approach Sec.~\ref{sec:NEGF}. The brief discussion on the different transport properties in terms of the transmission coefficient is also given. Then in Sec.~\ref{sec:results}, we show our results for the local density of states, transmission coefficient, electron conductance, thermal conductance and Seebeck coefficient. Finally in Sec.~\ref{sec:conclusion}, we conclude.

\section{Theoretical Background}
\label{sec:theo}

\subsection{Model}
\label{sec:model}
To study the electron transport of zigzag graphene nanoribbons, we consider the tight binding model which is an orthogonal $\pi$-orbital model. According to the latter model, the Hamiltonian of the system is described as
\bea
H_{0} &=& -t \sum_{\langle i,j\rangle, \sigma} \bigg[ c^{\dagger}_{i\sigma} c_{j \sigma} + c^{\dagger}_{j\sigma} c_{i \sigma}\bigg].
\label{eqn:M1}
\eea
Here $c_{i \sigma}$ ($c^{\dagger}_{i\sigma}$) is the electron annihilation (creation) operator having spin $\sigma$ on the site $i$, $t$ is the hopping integral between the nearest neighbor sites and controls the transfer of electron from one carbon atom to another carbon atom and $\langle i, j \rangle $ represents the summation over the neighboring sites. We have fixed the value of the hopping parameter $t=2.7$eV and assumed that $t$ is independent of external conditions.

\figpot
In this paper, our main focus is to study the transport corresponding to the confined electron within the quantum well. To do this analysis, we create a potential well with an application of an external electrostatic potential which is shown in Fig.~\ref{fig:pot}. The structure of this well is analogous to our quantum mechanics potential well problem. Mathematically, it can be represented as
\bea
V(x) &=& V_{0}\bigg[\Theta(x) - \Theta(x-L)\bigg],
\label{eqn:M2}
\eea
where $\Theta(x)$ is the Heaviside function and is defined as
\bea
\Theta(x) &=& \begin{cases}
               0,\hspace*{0.3cm}  x < 0 \\
               1, \hspace*{0.3cm} x \ge 0 
              \end{cases}
\label{eqn:M3}
\eea
and $V_{0}$ is a constant and its value is 0.5eV and $L$ is the length of the well and its value is 50nm. Using the definition of Heaviside function, the potential can be represented as\cite{surender_16}
\bea
V(x) &=& \begin{cases}
          V_{0},\hspace*{0.3cm} x <0 \hspace*{0.1cm}\text{and}\hspace*{0.1cm} x>L \\
          0,     \hspace*{0.5cm}   0 \le x \le L
         \end{cases}
\label{eqn:M4}         
\eea
Using Eqs.~(\ref{eqn:M1}) and (\ref{eqn:M4}), the total Hamiltonian in our case is described as
\bea
H &=& H_{0} + V(x).
\label{eqn:M5}
\eea
Considering this Hamiltonian, we will discuss the band structure, local density of states (LDOS) and electron transport properties of ZGNRs in later sections. Before going into the details, first we will give brief idea of the non equilibrium Green's function approach in the next subsection.
\subsection{Non-equilibrium Green function approach for electron transport}
\label{sec:NEGF}
The non equilibrium Green's function approach\cite{datta_book}, \cite{haug_book} is a powerful tool to study the dynamics of the mesoscopic system in which system is divided into three different regions namely left, center and right regions. In this work we are mainly interested in the electron dynamics in the center region of the system i.e. within the potential well. This latter can be analyzed by calculating the Green's function for the center region which includes the effects of the side left and right leads through the self energy contributions.

According to the this approach, the transmission coefficient is defined as\cite{datta_book}
\bea
T(E) &=& Tr\bigg(\Gamma_{L}G^{r}\Gamma_{R}G^{a} \bigg).
\label{eqn:NE1}
\eea
Here $\Gamma_{L,R}$ are the broadening functions and is defined as
\bea
\Gamma_{\alpha} &=& -2Im\Sigma_{\alpha}^{r},
\label{eqn:NE2}
\eea
where $\Sigma_{\alpha}^{r}$; $\alpha = (L,R)$ is the retarded self energy of the left and right leads of the system. $G^{r}$ ($G^{\alpha}$) is the retarded (advanced) Green's function and is defined as
\bea
G^{r} &=& \bigg( (E+i\eta)I - \mathcal{H} - \Sigma_{L}^{r} - \Sigma_{R}^{r} \bigg)^{-1}.
\label{eqn:NE3}
\eea
Here $E$ is the energy of an electron, $I$ is an identity matrix, $\mathcal{H}$ is the matrix representation of the Hamiltonian operator $H$, $\eta $ is infinitesimally small and it is introduced to remove the singularity and the self energies are defined as
\bea
\Sigma_{\alpha}^{r} &=& H_{C \alpha}G_{\alpha}^{0} H_{\alpha C}.
\label{eqn:NE4}
\eea
Here $H_{C \alpha}$ is the coupling between the center region and the side leads and $G_{\alpha}^{0}$ is the surface Green's function which is calculated by Lopez Sancho algorithm\cite{sancho_85}. Thus with the calculation of the retarded Green's function, the transmission coefficient can be calculated and the latter results can be used to calculate the electron transport properties.

Further these properties such as electron conductance, thermal conductance, Seebeck coefficient can be calculated by using Landauer-Buttiker formula\cite{landauer_57}, \cite{landauer_85} which is exact in the coherent transport limit\cite{datta_book}. It is a limit in which the inelastic scattering mechanisms are absent. Within this formalism, the quantities can be expressed in terms of the transmission coefficient as follows:
\begin{enumerate}
 \item Electronic conductance: It is a quantity which measures the flow of electronic current in a system and is expressed as\cite{esfarjani_06}
 \bea \nonumber
 G_{e} &=&  \frac{2q^{2}}{h} \int dE T(E) \bigg(-\frac{\partial f(E,\mu,T)}{\partial E}\bigg)\\
 &=& q^{2}L_{0}.
 \label{eqn:NE5}
 \eea
 
 \item Thermal conductance: It is a quantity which measures the flow of electron thermal current in a system and is expressed as\cite{esfarjani_06}
 \bea
 \sigma_{e} &=& \frac{1}{T} \bigg( L_{2} - \frac{L_{1}^{2}}{L_{0}} \bigg).
 \label{eqn:NE6}
 \eea
 \item Seebeck Coefficient: It is a quantity which measures the magnitude of the induced voltage in the response of the temperature gradient at two ends and is defined as\cite{esfarjani_06}
 \bea
 S &=& \frac{1}{qT} \frac{L_{1}}{L_{0}}.
 \label{eqn:NE7}
 \eea
\end{enumerate}
Here $L_{n}$ is defined as
\bea
L_{n} &=& \frac{2}{h} \int dE T(E) \big( E-\mu \big)^{n} \bigg(- \frac{\partial f(E, \mu , T)}{\partial E} \bigg),
\label{eqn:NE8}
\eea
where $\mu$ is the chemical potential, $q$ is the charge of electron, $T$ is the temperature, $f(E, \mu , T)$ is the Fermi-Dirac distribution function and represents as $f(E, \mu , T) = \bigg[\text{exp}\bigg({\frac{E-\mu}{k_{B}T}}\bigg) +1 \bigg]^{-1} $ and $k_{B}$ and $h$ are the Boltzmann and Planck's constant respectively.

\section{Results and Discussion}
\label{sec:results}
In this section, we present our results by changing the number of atoms per unit cells $N$, chemical potential $\mu$ and temperature $T$.

We focus in the confined region of the well and considered (N=4, 16, 64) atoms in our unit cell. Thus, the size of either of the interaction matrices or full Hamiltonian becomes N$\times$N (viz. 4$\times$ 4). Further on moving from one unit cell to next nearest unit cell, the wave function gains a phase shift. On moving towards the forward direction, there is $e^{ikx}$ phase shift in the wave function and the Hamiltonian can be written as $H_{01}$ which means the zeroth layer of unit cell is interacting with first unit cell times phase factor $e^{ikx}$. In a same fashion on backward direction, there is $e^{-ikx}$ phase shift and the Hamiltonian becomes $H_{10}$ times  $e^{-ikx}$\cite{surender_17}. Hence the total Hamiltonian which consists three parts becomes 
\begin{equation}
 H=H_{00}+H_{01}e^{ikx}+H_{10}e^{-ikx},
\end{equation}
where $H_{00}$ is the zeroth layer of unit cell interacting with itself\cite{surender_16}. Order of this Hamiltonian depends upon the number of atoms in the unit cell. Since we know from the simple quantum mechanical concepts that solution inside the finite potential well are sin(kx) or cos(kx), same thing is happening here in our work.

In the matrix form, the total Hamiltonian can be written in the tri-diagonal form and forms the sparse matrix. It is expressed as 
\begin{equation}
 [H] = \left( \begin{array}{cccc}
 [H_{00}] & [H_{01}]  & \dots & [0] \\

 [H_{10}]    & [H_{00}] & \dots & \dots \\
 
 \dots & \dots & \dots & \dots \\
 
 \dots & \dots & \dots & [H_{01}] \\
 
 [0]   & \dots & [H_{10}] & [H_{00}] \end{array} \right)
\end{equation}
By diagonalizing this Hamiltonian and plotting in terms of k will give us bandstructure. This can be seen in Fig.~\ref{fig:BS}.

\figBS

Using this Hamiltonian, we have analyzed various quantities in the following subsections.
 \subsection{Density of states}
 \figLDOS
 The density of states is defined as the imaginary part of the trace of the Green's function with multiplication of 1/$\pi$ factor i.e.
 \begin{equation}
  \text{DOS} = -\frac{\text{Im}[Tr(G)]}{\pi}.
 \end{equation}
Using this definition and Eq.~(\ref{eqn:NE3}), we plot the density of states in Fig.~\ref{fig:LDOS}. Here we show DOS for different number of atoms per unit cell i.e. $N=4$, $16$ and $64$, keeping other parameters fixed. We find that around $E=0$, the density of states show the occurrence of peak due to the formation of edge states\cite{li_08}, \cite{tao_11}. It is also observe that LDOS away from $E=0$ increases with the increase of $E$. This is a signature of the more transmission of the electrons from the potential well. Further with the increase of $N$, the fluctuations in LDOS near the Fermi energy increases as shown in Fig.~\ref{fig:LDOS}.
 \subsection{Transmission Coefficient} 
 \figTCo
 Figure~\ref{fig:TCo} shows the electronic transmission coefficient as a function of energy in the case of zero bias. This has been done using Eqs.~(\ref{eqn:NE1}) and (7). In this figure, we find that the transmission spectra shows a dip around the Fermi energy $\epsilon_{F}$ which gives the signature of the band gap between the conduction and the valence bands. We also find that $T(E)$ remains unity between the energy range $-2$eV to $2$eV in Fig.~\ref{fig:TCo4} corresponds to the case for number of atoms per unit cell $N=4$. If we move towards Fig.~\ref{fig:TCo16} for $N=16$, we find that the energy range for the unity value of the transmission spectra decreases and it further decreases more for $N=64$. This indicates that there will be no current flow within this energy range of the system\cite{zhang_08}. At high energies, electrons start tunnel across the energy barrier, resulting the more transmission coefficient. One more important feature of $T(E)$ is its behavior in the steps. This implies that 
the electrons can pass through the center region of the quantum well without undergoing any scattering mechanisms such as electron-phonon, electron-impurity, phonon-phonon. These mechanisms are important to understand the temperature behavior of the transmission spectra which is beyond the scope of the present article. 
 \subsection{Electronic Conductance} 
 \figEC
 In Fig.~\ref{fig:EC}, we have plotted the electronic conductance $G_{e}$($G_{0}$) as a function of temperature $T$(eV) by considering different number of atoms per unit cell such as 4, 16, and 64. Here we show the variation in electronic conductance at different chemical potentials $\mu$(eV)$=$0.0, 0.5, 0.7, 1.0. It is found that $G_{e}$ is constant throughout the temperature regime for $\mu = 0.5$, $0.7$ and $1.0$eV as shown in Fig.~\ref{fig:EC}. However for $\mu=0.0$, it first increase with the increase in the temperature $T$ and then saturates at high temperature as shown in Fig.~\ref{fig:EC4}. This scenario can be understood with the reference of the Fermi energy level. In the case of zero chemical potential i.e. $\mu=0.0$, the Fermi energy lies in between the valance and conduction subbands. Thus only few electrons can excite to the conduction subband and lead to small value of electronic conductance. While for the case of $\mu = 0.5$, $0.7$ and $1.0$, due to the shift of the Fermi energy 
level into the conduction subband, the electrons can easily excite into the conduction subband and lead to more value of the electronic conductance. Similar picture can also be seen in Figs.~\ref{fig:EC16} and \ref{fig:EC64} corresponds to the number of atoms per unit cell $N=16$ and $64$ respectively. Here the difference is that the magnitude of the electronic conductance is more than the case for $N=4$. This is due to the consideration of large value of $N$. Here the another different feature is the non saturation behavior of $G_{e}$ at high temperature.
 \subsection{Thermal Conductance} 
 \figTC
 Next, we present the ballistic electronic thermal conductance as a function of temperature in Fig.~\ref{fig:TC}. Here we observe that with the shift in the Fermi energy level by increasing the chemical potential, the thermal conductance increases with increasing temperature. In the low temperature regime, the magnitude of $\sigma_{e}$ corresponds to the large shift in the Fermi energy level is high as compared to small shift i.e. to the case of small $\mu$. Another feature in the behavior of the electronic thermal conductance is that it shows crossover depending on the value of $N$ at high temperature value. This can be observed at $T=0.6$, $0.4$ and $0.3$ in Figs.~\ref{fig:TC4}, \ref{fig:TC16} and \ref{fig:TC64} corresponds to $N=4$, $16$ and $64$ respectively. This crossover due to the scattering caused by the confined electrons in the middle region of the potential well from the edges. Further at high temperature, the behavior of the electronic thermal conductance is opposite to that at low 
temperature due to not taken into account the phonon contribution which play an important role at high temperature regime. These results in the low temperature regime are in accord with the results existed in the literature\cite{saito_06}.
 \subsection{Seebeck Coefficient} 
 \figSC
In Fig.~\ref{fig:SC}, we present the thermoelectric coefficient, named as Seebeck coefficient using Eq.~(\ref{eqn:NE7}) for different number of atoms per unit cells and chemical potential. Here we observe that at $\mu=0.0$, the Seebeck coefficient is zero due to the symmetric valence and conduction subbands. At higher chemical potential, it increases with the temperature and then peaks at some temperature value and then decays. This behavior is due to the weighted factor $(E-\mu)$ in the integrand of the numerator of Eq.~(\ref{eqn:NE7}). When this weighted factor becomes maximized i.e. the chemical potential becomes of the several $k_{B}T$, the Seebeck coefficient gives maximum value\cite{ouyang_09}. This can be seen in Fig.~\ref{fig:SC}. Similar to the case of the electronic and thermal conductance, the magnitude of the Seebeck coefficient also increases with the increase of $N$. As an example the value of Seebeck coefficient corresponds to $\mu=1.0$(shown in pink color) is around 0.12 at T=0.6 in the case 
of $N=4$ (Fig.~\ref{fig:SC4}). As we move to Fig.~\ref{fig:SC16} corresponds to $N=16$, it becomes roughly $1.0$ and then $1.1$ for $N=64$ in Fig.~\ref{fig:SC64}.  

\subsection{Figure of Merit}
  \figFM
The figure of merit is the ratio of the thermoelectric properties and is an important quantity to find the efficiency of the material. It can be defined using Eqs.~(\ref{eqn:NE5}),(\ref{eqn:NE6}) and (\ref{eqn:NE7}) as
\begin{eqnarray} \nonumber
 ZT &=& \frac{S^{2}G_{e}T}{\sigma_{e}}\\
 &=& L_{1}^{2} \bigg( L_{2} - \frac{L_{1}^{2}}{L_{0}} \bigg).
\end{eqnarray}
Using Eq.~(\ref{eqn:NE8}, we have plotted the figure of merit i.e. $Z$ as a function of temperature at different chemical potential $\mu$ and $N$ in Fig.~\ref{fig:FM}. Here we observe that at very low temperature, there is vanishing figure of merit. This corresponds to the behavior of thermal conductance and Seebeck coefficient below the crossover regime i.e. below $0.4$ eV. Further above this temperature, it increases and also increases with the chemical potential. At high temperature, it decreases. However with the decrease of number of atoms $N$, the magnitude of $ZT$ decreases. It indicates that the electron contribution to the figure of merit is important at low temperature scale and less pronounced at high temperature regime where phonon play their major role\cite{jiang_11}. We also find that the efficiency of the zigzag graphene nanoribbons can be increased with the increase of the temperature as shown in Fig.~\ref{fig:FM}.

\section{Conclusion}
\label{sec:conclusion}
We have studied the transport properties of the zigzag graphene nanoribbons within the non-equilibrium Green's function technique. Transmission is clearly positive in the confined region of the well and it indicates metallic behavior. When we are increasing number of atoms, number of plateaus are increasing respectively of the order of quantum of conductance. We have done calculation for multi subbands in this work. 
\par We have studied the effects of the variation of the number of atoms per unit cell and the chemical potential. In the whole analysis, the electronic contribution to the transport properties is investigated. We find that the energy range for the transmission coefficient to be remain unity and decreases with the increase of more confined electrons within the potential well. It is found that the electrons contribute to the figure of merit at low temperature regime. Also the figure of merit increases with the number of atoms due to the more magnitude of the thermal conductance and the Seebeck coefficient. In addition to these, we also find that the increase in the value of chemical potential enhance the magnitude of transport properties which further enhance the figure of merit. Here we remark that we have applied NEGF approach of ballistic transport to investigate the transport properties. At high 
temperature, the phonon contribution should play their role which lead to enhance the figure of merit in this temperature regime.

\end{document}